\documentclass[pre,twocolumn,nofootinbib]{revtex4-1}
\usepackage{graphicx}
\usepackage{color}
\usepackage{epsfig}

\begin{document}

\title{Excess Vibrational Density of States and the Brittle to Ductile Transition
in Crystalline and Amorphous Solids}

\author{Jeetu S Babu${^1}$}
\author{Chandana Mondal$^{2}$}
\author{Surajit Sengupta$^{1}$}
\author{Smarajit Karmakar$^{1}$}
\email{smarajit@tifrh.res.in}
\affiliation{$^1$ Centre for Interdisciplinary Sciences, Tata Institute of
Fundamental Research, 21 Brundavan Colony, Narisingi, Hyderabad, India, $^2$ Department of Chemical Physics, Weizmann Institute of Science, Rehovot, Israel}

\date{\today}

\begin{abstract}
The conditions which determine whether a material behaves in a brittle or ductile fashion on mechanical loading are still elusive and comprise a topic of active research among materials physicists and engineers. In this study, we present results of {\em in silico} mechanical deformation experiments from two very different model solids in two and three dimensions. The first consists of particles interacting with isotropic potentials and the other has strongly direction dependent interactions. We show that in both cases, the excess vibrational density of states is the fundamental quantity which characterises the ductility of the material. Our results can be checked using careful experiments on colloidal solids.    
\end{abstract} 

\maketitle

\section{\label{sec:level1} Introduction:}
Understanding the mechanical behaviour of glassy materials~\cite{09Cav,debenedetti,kob,kobAndersenPRL, 
sastry,06BBMR,08BBCGV,09KDS,11BB,14KDS} has engaged the attention of materials scientists because of both its technological ramifications and scientific interest~\cite{BMGintroMRS,johnsonMRS,2011Dauchot, GreerNatMat, GreerMSER}. Glasses, in some ways represent an extreme limit of a supercooled liquid whose viscosity has increased dramatically eg.  by almost $14$ orders of magnitude with decreasing temperature within a narrow range, ($\sim 100K$) without any major change in structure reaching a value of $\approx 10^{13}$Poise \cite{weeksscience, edigerdynamic}. While the question of why viscosity rises so rapidly has remained elusive~\cite{ritortsollich,lubchenkowolynes,RFOT2,biroliBouchaudRFOT,11BB}, in this paper, we investigate the consequence of this ultra slow relaxation on the failure properties of amorphous materials under external load. 

\begin{figure}[!h]
\centering
\vskip +0.2cm
\includegraphics[width=0.4\textwidth, angle =0]{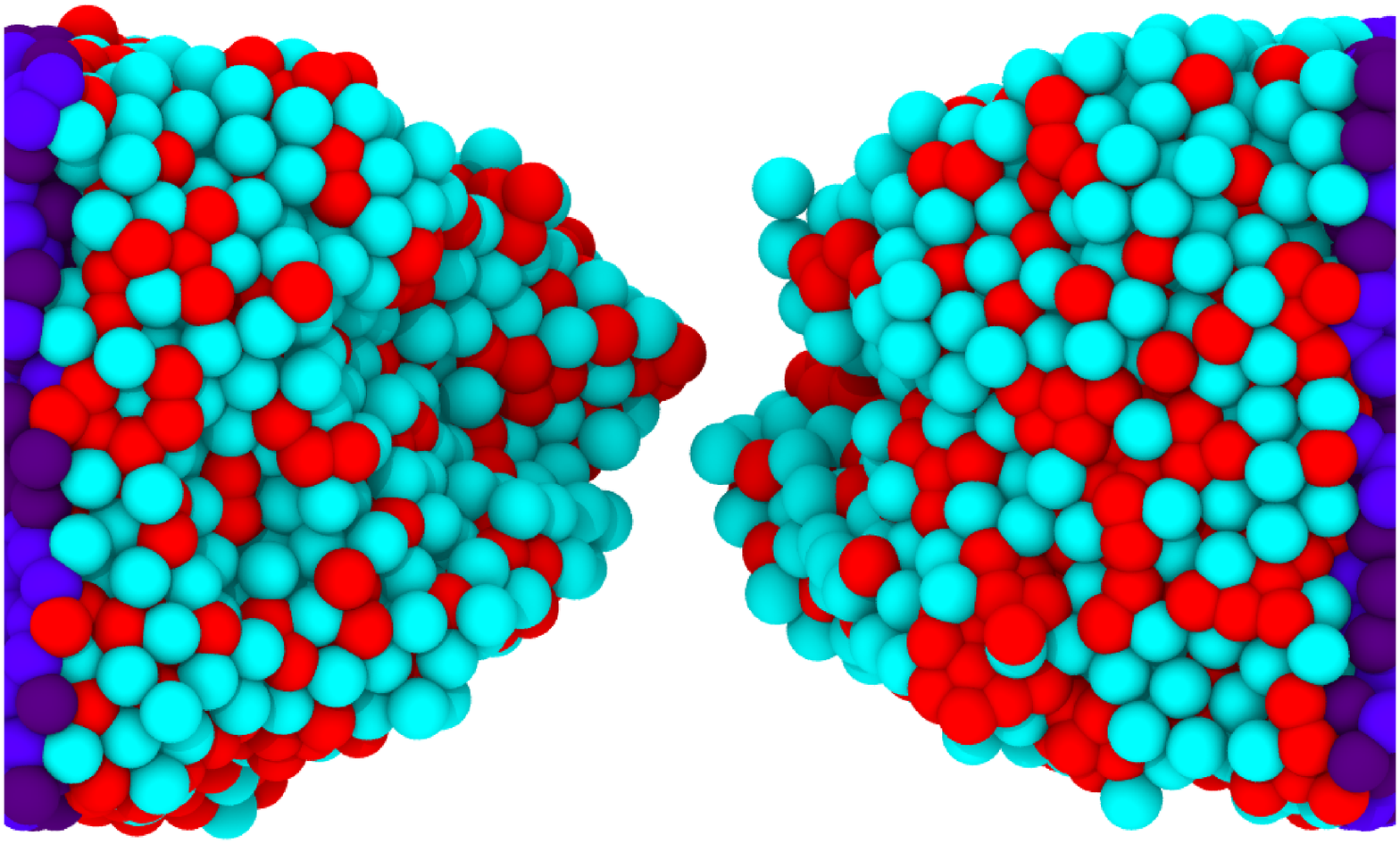}\\
\includegraphics[width=0.45\textwidth, angle = 0]{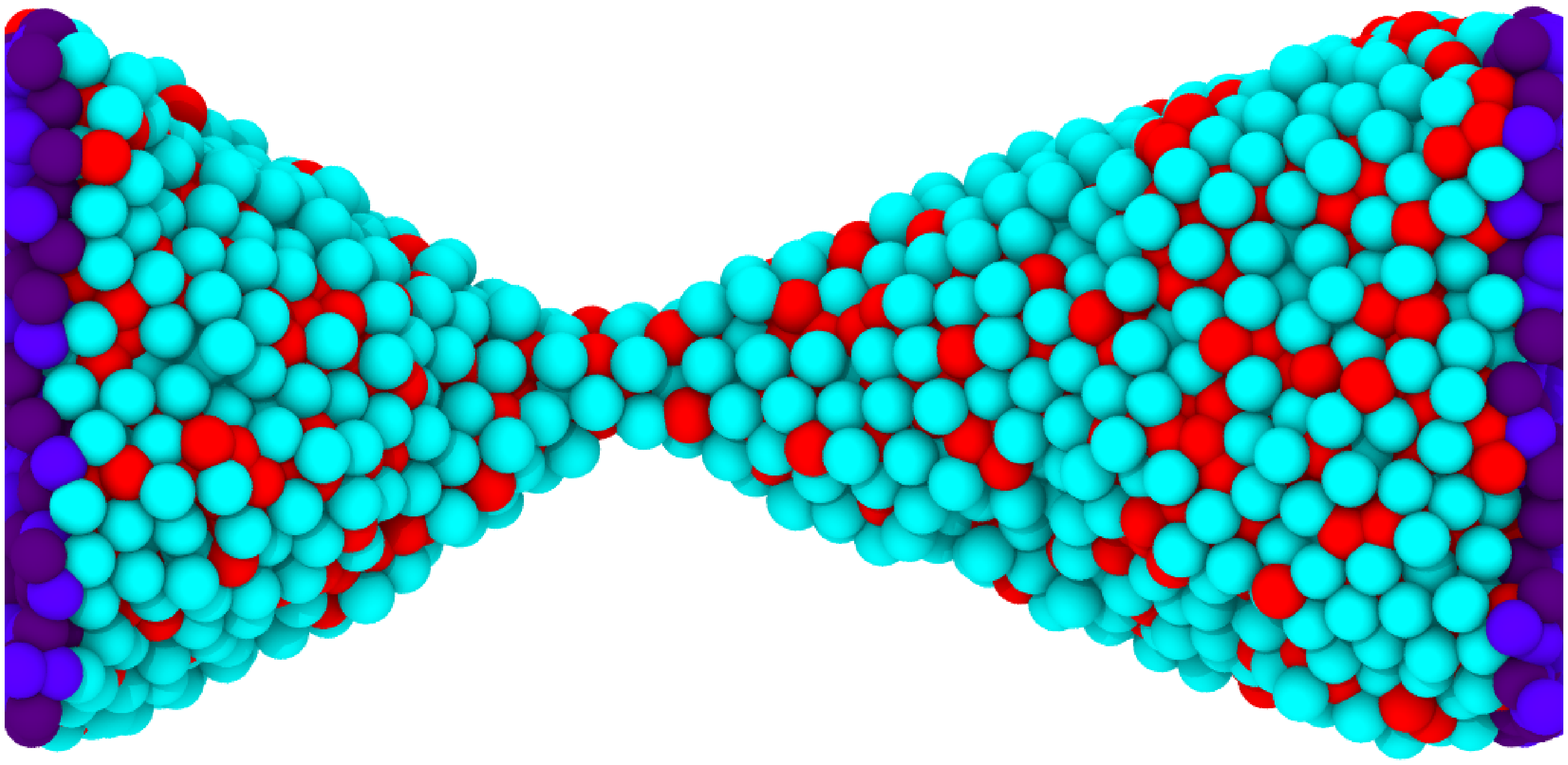}
\caption{Snapshot picture of a brittle ( top panel ) and a ductile (bottom panel) 
materials at the point of fracture. Note the drastic difference between the fracture
morphologies for these two cases.}
\label{fig:1}
\end{figure}

In recent experiments it was found that glassy materials can have much larger yield stress compared to their crystalline counterparts with similar compositions, but they fail 
catastrophically under external loading\cite{GreerNatMat, GreerMSER}. Thus one of 
the major questions which needs to be addressed in order to be able to use glassy 
materials for industrial applications, is how much plastic deformation can the 
material withstand before it fails or in other words will the material experience 
brittle fracture or ductile fracture \cite{GreerNatMat2006}. 
In Fig.\ref{fig:1}, typical failure events are shown for a brittle ( top panel ) 
and a ductile material ( bottom panel ). One can clearly see that the morphology
of the fracture is very different for the two cases. For the brittle case one has 
somewhat rough fracture surface, on the other hand for the ductile fracture, the 
system forms a neck before breaking completely. It is believed that for brittle 
materials cavities form during tension and the different cavities eventually 
percolate to form the brittle crack, on the other hand for the ductile materials 
the failure mechanism does not involve cavities\cite{ZylbergPRE2013}. It is important 
to mention that the microscopic details of this cavitation is still not clearly 
understood nor the necking behaviour in ductile materials. 

Recent molecular dynamics 
(MD) simulations have investigated the intrinsic relation between the properties 
of the inter-particle potential in order to understand the microscopic origin of 
brittle to ductile transition in amorphous solids \cite{2011Dauchot}. It was shown 
that the degree of ductility is intrinsically related to the growing importance 
of the contribution of plastic modes to the vibrational density of states of the 
material. Similarly Mizuno et al.\cite{2013Mizuno} and Goodrich et al.
\cite{2014Goodrich} have also observed that the mechanical and thermal properties 
of a solid are directly related to the density of normal modes of vibration. In 
\cite{2014Goodrich}, it is found that the physics of crystalline solids with 
defects are better described by the physics of amorphous solids near jamming 
transition if the concentration of defects are slightly large but not large enough
to completely destroy the bulk crystalline order in the system. It was shown that
vibrational density of states for these crystalline solids with defects are very
similar to the amorphous solids near jamming transition. Their main conclusion 
is that jamming point of an amorphous solids is probably the better reference point
than an ideal crystalline state to understand the physics of crystalline solids 
with defects beyond a small critical defects concentration.     

The aim of this paper is to investigate the connection between the mechanical and 
vibrational properties of both crystalline and amorphous solids in an unified 
framework and try to come up with a measure which can characterises these properties 
of the system irrespective whether the material is crystalline or amorphous in 
nature. This is performed via extensive Molecular Dynamics Simulation by studying 
different systems in which the mechanical behavior of the material is varied 
continuously by tuning a control parameter like the interaction range 
of the inter-atomic potential, size  disorder of different particles in the system and 
the quench dynamics of a colloidal system.     

The structure of the paper is as follows: In Sec.~\ref{sec:model} we will describe the 
model systems studied, the details of the parameters of the model and the how the 
simulations are done for all these model systems. The numerical 
experiments and their results are described in Sec.~\ref{sec:RESULTS} and 
finally, the main conclusions and future directions will be discussed in 
Sec.~\ref{sec:conclusions}.

\section{\label{sec:model} Models and Simulations Details:}
To show that our results are generic, we report studies on two very different model solids. The first one consists of particles interacting with isotropic potentials and is a more or less accurate representation of a metallic glass. We study this model in both two and three dimensions. The bulk of the results presented in this manuscript relates to this model. We also present results for a second model solid which is composed of particles with strongly direction dependent interactions. This solid can be used to mimic gels, silicates and other such anisotropic materials. We describe each of these models in somewhat more detail below.

\subsection{Model A:}
The model studied is a generic glass former in both two and three dimensions. 
It is a binary mixture whose amount of bi-dispersity of small and large particles 
was chosen to avoid any crystallization. The particles interact by an inter-particle 
potential given by
\begin{equation}
\label{eqn:2} 
\phi \left ( \frac{r_{ij}}{\sigma _{ij}} \right )=\left \{  
\begin{array}{lr}
4\epsilon \left [ \left ( \frac{\sigma _{ij}}{r_{ij}} \right )^{12}- \left ( 
\frac{\sigma _{ij}}{r_{ij}} \right )^{6}\right ],\quad r_{ij}/\sigma_{ij}\leq 
r_{min} \\ 
\epsilon \left [ a\left ( \frac{\sigma _{ij}}{r_{ij}} \right )^{12}- 
b\left(\frac{\sigma _{ij}}{r_{ij}} \right)^{6} \right.\\
+ \left.\sum_{l=0}^{2}c_{2l}\left ( \frac{r_{ij}}{\sigma _{ij}}\right )^{2l} \right ], \; 
r_{min} < r_{ij}/\sigma_{ij}< r_{co}\\ 
0, \quad r_{ij}/\sigma_{ij}\geq r_{co}
\end{array}
 \right.
\end{equation}
where $r_{min}$ is the length where the potential attains its 
minimum, and $r_{co}$ is the cut-off length for which the potential 
vanishes. The coefficients $a$, $b$, and $c_{2l}$ are chosen such that the 
repulsive and attractive parts of the potential are continuous with two continuous 
derivatives at the potential minimum and the potential goes to zero 
continuously at $r_{co}$ with two continuous derivatives as well. 
$\epsilon$ is the unit of energy, and $k_{B} = 1$.

We have performed simulations where the cut off distance $r_{co}$ was varied 
from $1.2$ to $2.2$ keeping all other parameters of the interaction potential 
unchanged. The interaction length scale $\sigma_{ij}$ between any two small 
A particles is $\sigma_{AA} = 1.0$ and similarly for between one small A and 
one large B particles is $\sigma_{AB} = 1.18$. The interaction length scale
between two large B particles is $\sigma_{BB} = 1.4$.

We have also done simulations with $\sigma_{ij}$ varied so that the system 
traverses from a crystalline state to an amorphous state. This is known as 
amorphization transition \cite{2013Mizuno}. Here, an effective interaction 
length scale $\sigma_{eff}$ is maintained such that $\sigma_{eff}=\sum_{ij} 
x_{i}x_{j}\sigma_{ij}^{3}$ , where $x_{1} = x_{2} = 1/2 $ 
are the fractions of the two components. $\lambda = \frac{\sigma_{11}}
{\sigma_{22}}\leq 1$ measures the degree of bi-dispersity in the system, 
$\sigma_{11}$ and $\sigma_{22}$ are determined using the condition that 
$\sigma_{eff}$ is the same as that for original model described before. 
We kept $r_{co} = 1.60\sigma_{ij}$. The simulations are done in both two 
and three dimensions to study the effect of dimensionality in the results
reported.

NVT MD simulations are done in a cubic simulation box with 
periodic boundary conditions in both two and three dimensions for all the 
model systems using LAMMPS package and visualization of the MD trajectory is 
done using VMD open source software \cite{1995Plimpton,1996Humphrey}. We use 
the modified leap-frog algorithm with the Berendsen thermostat to keep the 
temperature constant in the simulation runs as implemented in LAMMPS. Any
other thermostat does not change the results quantitatively as we are mostly
interested in configurational changes in the system instead of momentum 
correlations. Length, energy and time scales are measured in units of 
$\sigma_{AA}$, $\epsilon_{AA}$ and $\sqrt{\sigma_{AA}^2/\epsilon_{AA}}$.  
The integration time steps used is $dt = 0.005$ in the studied temperature 
range. The number of particles used for is $N = 2000$ and the bi-dispersity 
ratio was $50:50$.
\begin{figure}[!h]
\begin{center}
\vskip -0.9cm
\includegraphics[scale=0.25]{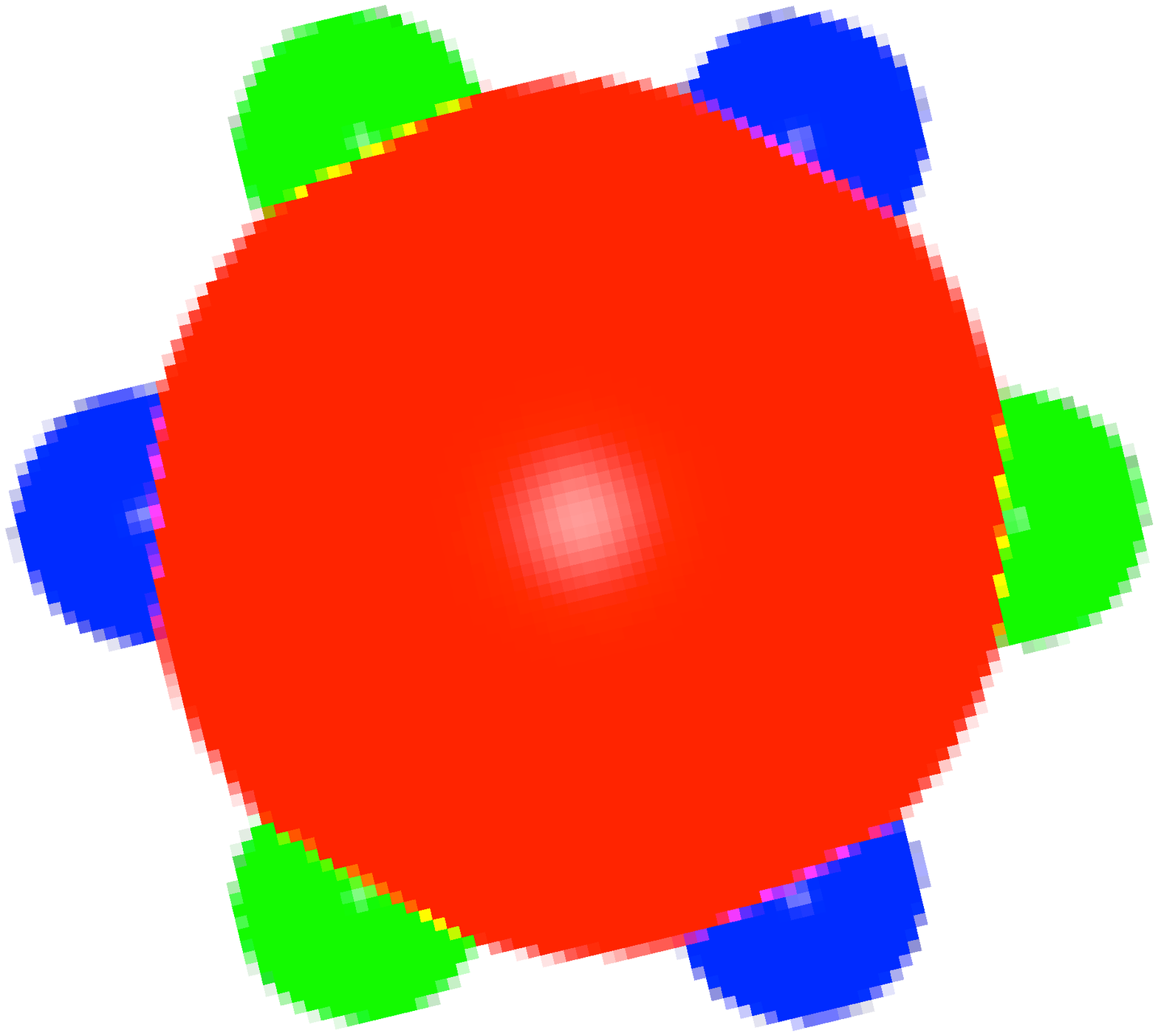}
\end{center}
\vskip -0.9cm
\caption{The schematic picture of a patchy spherical colloidal particle (red circle) 
with patches (blue and green semi-circles) along the equator.}
\label{patchy}
\end{figure}

\subsection{Model B:}
While Model A, described above, has isotropic interactions, in order to study 
whether the results reported here are generic for all glass forming 
liquids, we also perform simulations with a model solid with strongly anisotropic interactions viz. a model colloidal solid with patchy interactions in two dimensions. The details of the potential modelling the patchy colloid is based on the angle-dependent interactions of Hamaker\cite{Hamaker1937,
2003Everaers} and Lennard-Jones (LJ) type as implemented in the MD simulation package 
LAMMPS\cite{1995Plimpton}. In our model, each molecule consists of one central, 
large, spherical particle with six small equidistant patches of alternating 
types on it's equator. In Fig.\ref{patchy}, we draw a schematic for the patchy 
colloid particle. We refer to the central particle shown as a large red circle 
as a $Type-1$ particle and the blue and green semicircular  patches on it's 
equator as $Type-2$ and $Type-3$ particles respectively.  
  
The interaction between two $Type-1$ particles is a Hamaker interaction 
\cite{Hamaker1937} between two large size colloid particles. The interaction 
between a $type-1$ particle and a $type-2/type-3$ particle is the interaction 
between a large size colloid particle and a solvent LJ particle. Two $type-3$ 
particles or a $type-2$ particle and a $type-3$ particle interact with simple 
LJ interaction. Details of the interactions are given in the appendix of 
\cite{15MKS}. The sizes and interaction strengths between $type-1$, $type-2$ 
and $type-3$ particles are also tabulated in the appendix of \cite{15MKS}. 
All quantities in the table are expressed in reduced units. The unit of length 
and energy are $\sigma_{11}$ and $A_{11}$ respectively. We choose the mass of 
each molecule $\sigma_{11}=A_{11}=m=1$ without loss of generality, where $m$ is 
the mass of each particle.

We carry out molecular dynamics (MD) simulations on this system
with an integration time-step $\delta t = 5\times10^{-3}$ in constant $NAT$ ensemble, 
where $N=864$ is the number of patchy-colloids and $A$ is the area in two dimensions. 
The system-temperature is kept fixed at the desired value using dissipative 
particles dynamics (DPD) thermostat\cite{dpd-LAMMPS} as implemented in LAMMPS.

\subsection{Deformation protocols:}
In all these systems the numerical experiments are designed as follows. We 
first equilibrate the model liquids at some high temperature and then cool
it to low temperature below the experimental glass transition temperature
defined as the temperature where the relaxation time becomes $10^6$. Then
at that low temperature we perform constant pressure and temperature(NPT) 
simulations at zero pressure such that one can now remove the periodic boundary 
condition. The initial simulations at high temperature and high pressure was 
done for $5\times10^6$ time steps, then the system temperature was gradually 
reduced in $5\times10^5$ steps, and finally the NPT simulation at zero 
pressure was run for $5\times10^5$ steps.  We now define two side walls  
at the two ends of the solid in the $x$ direction by pinning the particles in 
the end region. The typical size of this wall is around three inter-particle
diameter. The other boundaries are made free. Next the walls are moved by an 
increment equal in size and opposite in sign, i.e., the system is subjected 
to uniaxial strain. If the material is brittle it cannot deform much before it 
fractures, while for a ductile substance the distance between the walls 
increases almost by a factor of two before a thin neck forms due to the plastic 
deformations and eventually breaks. To quantify ductility of the system we 
use the percent of elongation of the system before breaking as given by
\begin{equation}
\label{definition} 
\Gamma =\left ( \frac{L_{f}-L_{0}}{L_{0}}\right )\times 100
\end{equation}
where, $L_{0}$ is the initial (before pulling) length of the system along the 
tension direction ( $x$ direction ) and $L_{f}$ is the final length before the 
system breaks into two parts.

\section{\label{sec:RESULTS} RESULTS}
In this section we present results from our simulations following the protocols described for each of the models. The bulk of our results are for the solid described by Model A. We show later that our main finding, i.e. the close relationship between the vibrational density of states and fracture behaviour of solids is also borne out in Model B. 

\subsection{Model A:}
\begin{figure}[!t]
\includegraphics[width=0.45\textwidth]{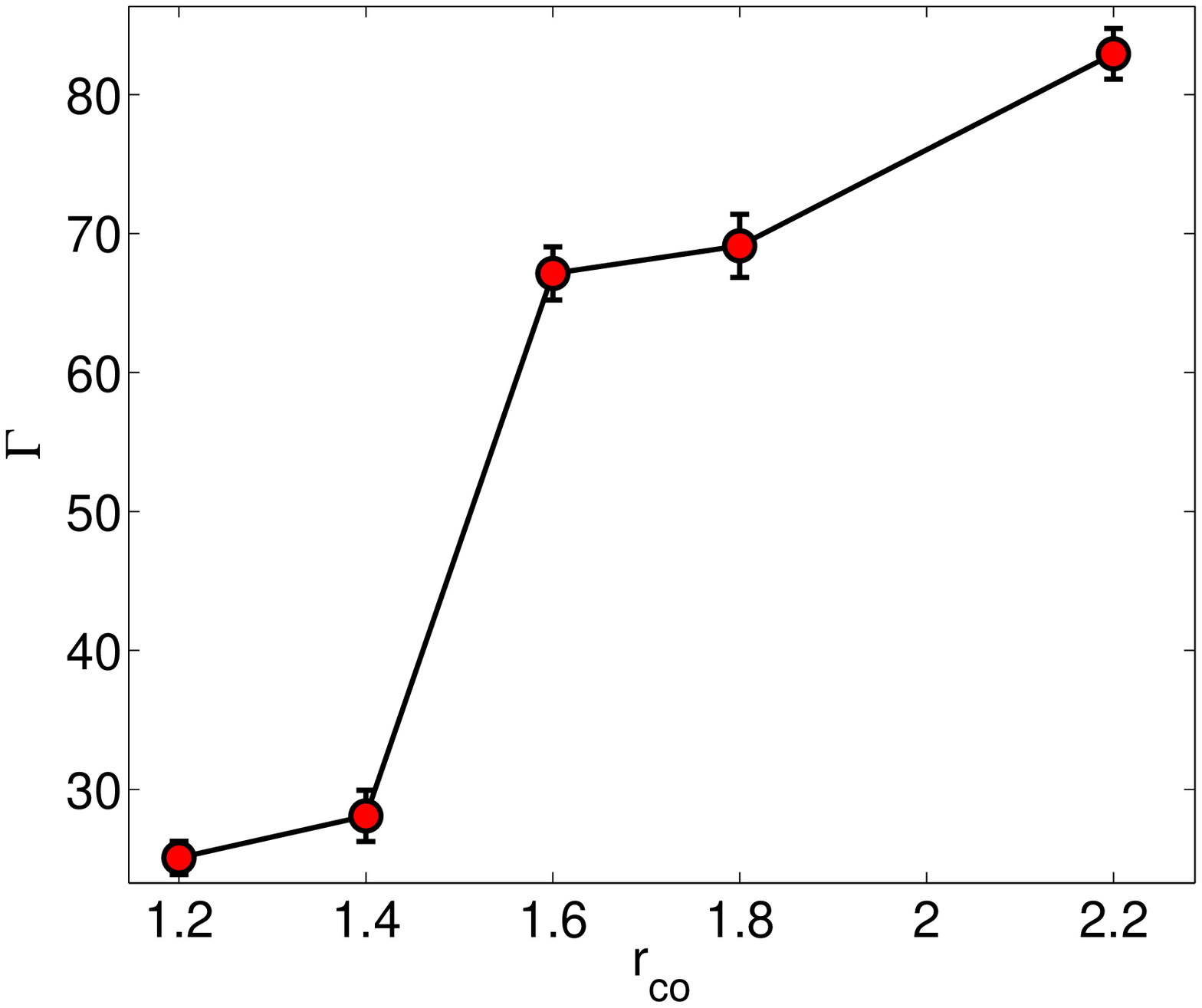}
\includegraphics[width=0.46\textwidth]{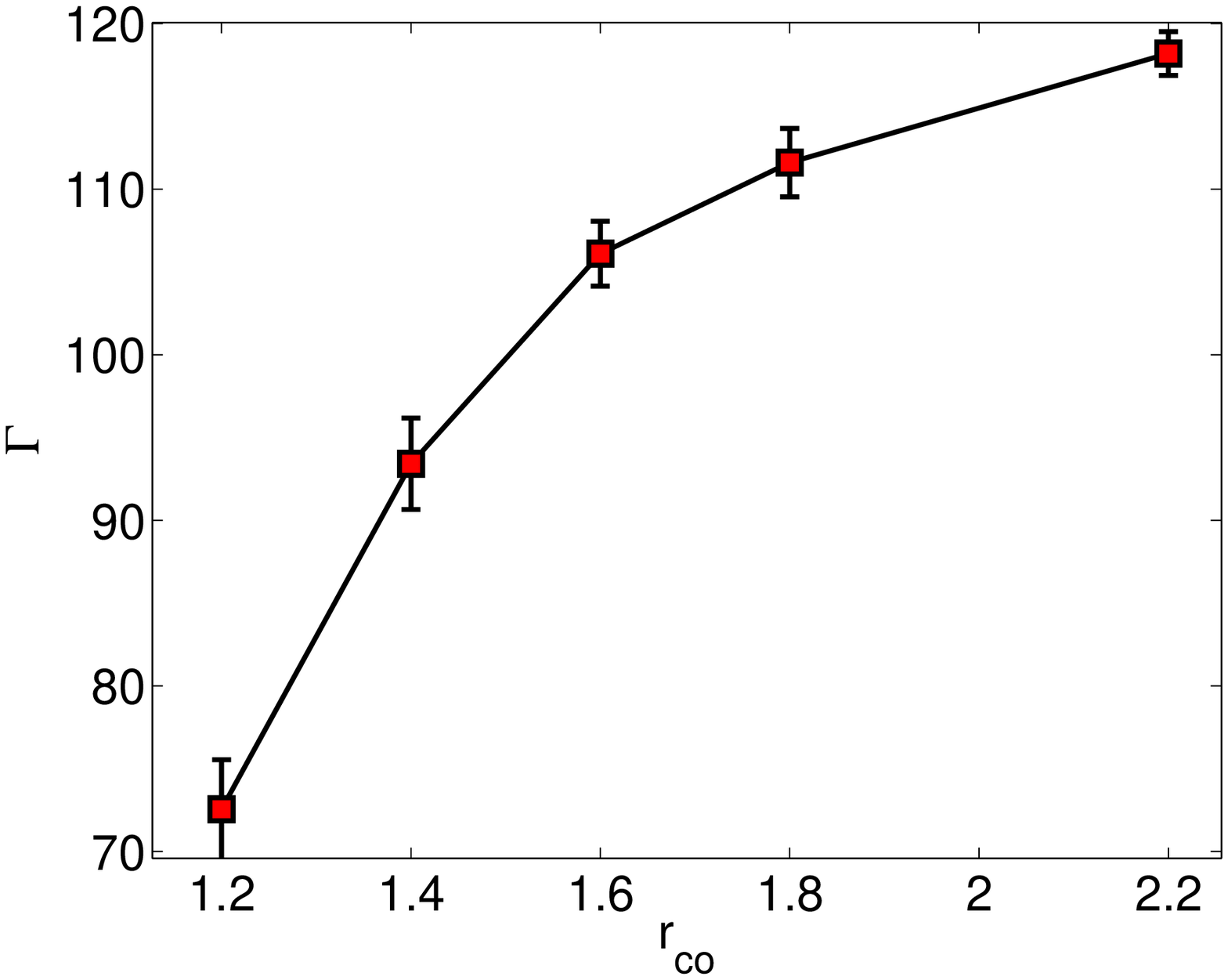}
\caption{Top panel: Variation of ductility parameter $\Gamma$ (see text for the
definition) with cut-off range for a model two dimensional glass forming liquid.
$\Gamma$ changes somewhat sharply as the interaction range $r_{co}$ is increased
and seems to reach a plateau once $r_{co}$ becomes large enough to include the 
second neighbour shell in the pair correlation function (see text for details). 
Bottom panel: similar analysis done for the same model system in three dimensions.
}
\label{fig:2}
\end{figure} 

In Fig. \ref{fig:1} the snapshots of a brittle (top panel) and ductile material
(bottom panel) under uniaxial strain, clearly show that brittle and ductile 
materials respond to external strain very differently. In the case of a brittle 
material, the material cannot withstand strain for a large deformation and hence 
it fails quickly while for a ductile material there is a neck formation which helps it to withstand strain for a longer 
period. This behavior can be quantified using the parameter $\Gamma$ as described in 
the previous section. In the upper panel of Fig. \ref{fig:2}, we show the dependence
of $\Gamma$ as a function of increasing range of interaction of the inter-particle
potential, $r_c$ for the two dimensional glass forming liquids. One can clearly
see that the ductility parameter modeled by $\Gamma$ is somewhat small and does not
increase much until the interaction range starts to include the second neighbours. 
The first neighbour shell is up to a distance $1.4$ in units of inter-particle diameter
as obtained from the dip of pair correlation function, $g(r)$ after the first peak. 
$\Gamma$ then somewhat sharply increases once $r_c$ is increased beyond the first 
interaction shell and then tend to saturate once the interaction range increases 
beyond the second neighbour distance. Similar behaviour is observed for the
three dimensional system as shown in the lower panel of Fig.\ref{fig:2} although
the increase in $\Gamma$ is somewhat smoother than two dimensional system. 

\begin{figure}[!h]
\centering
\includegraphics[width=0.5\textwidth]{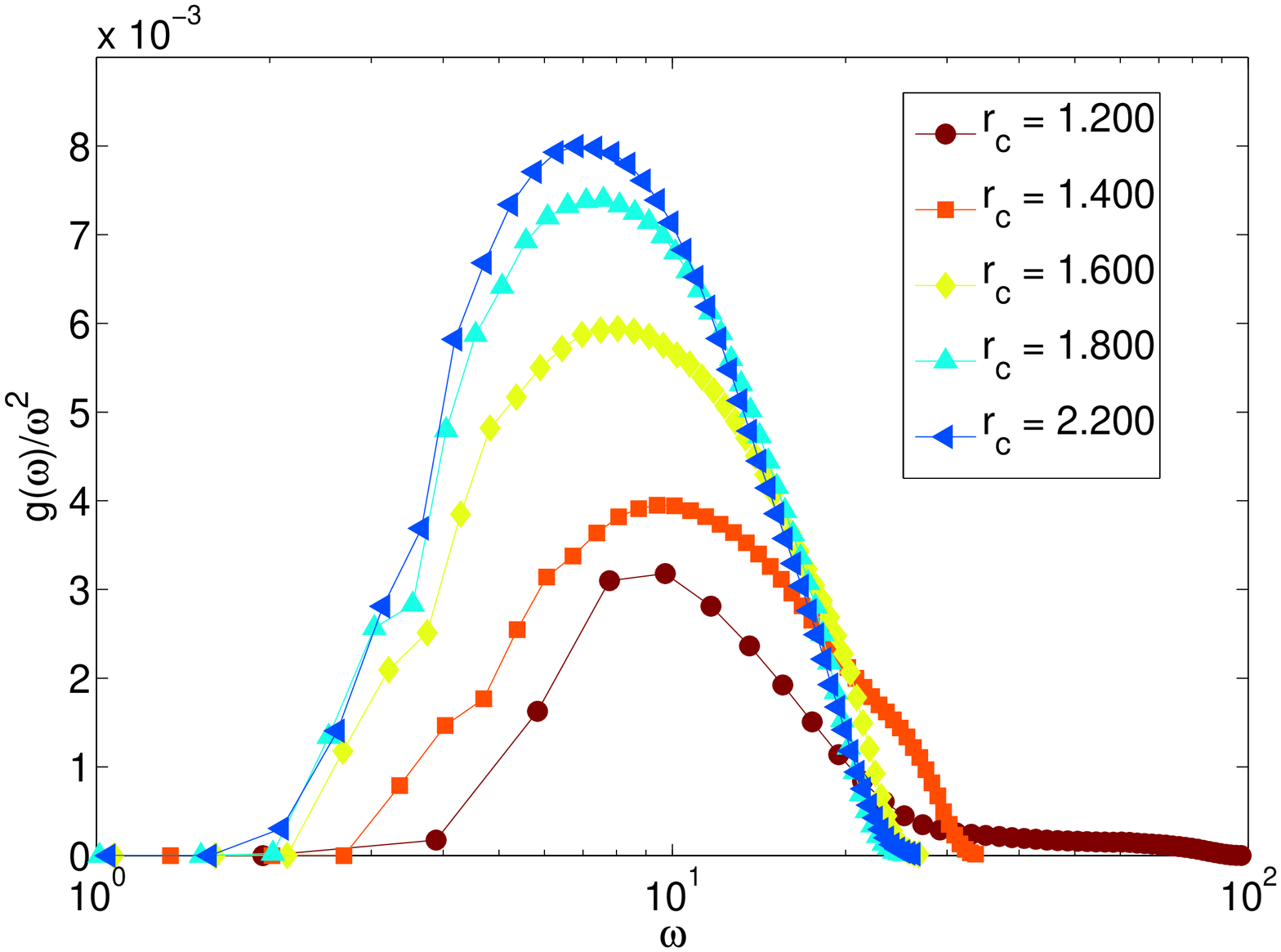}\hskip -0.2cm
\includegraphics[width=0.5\textwidth]{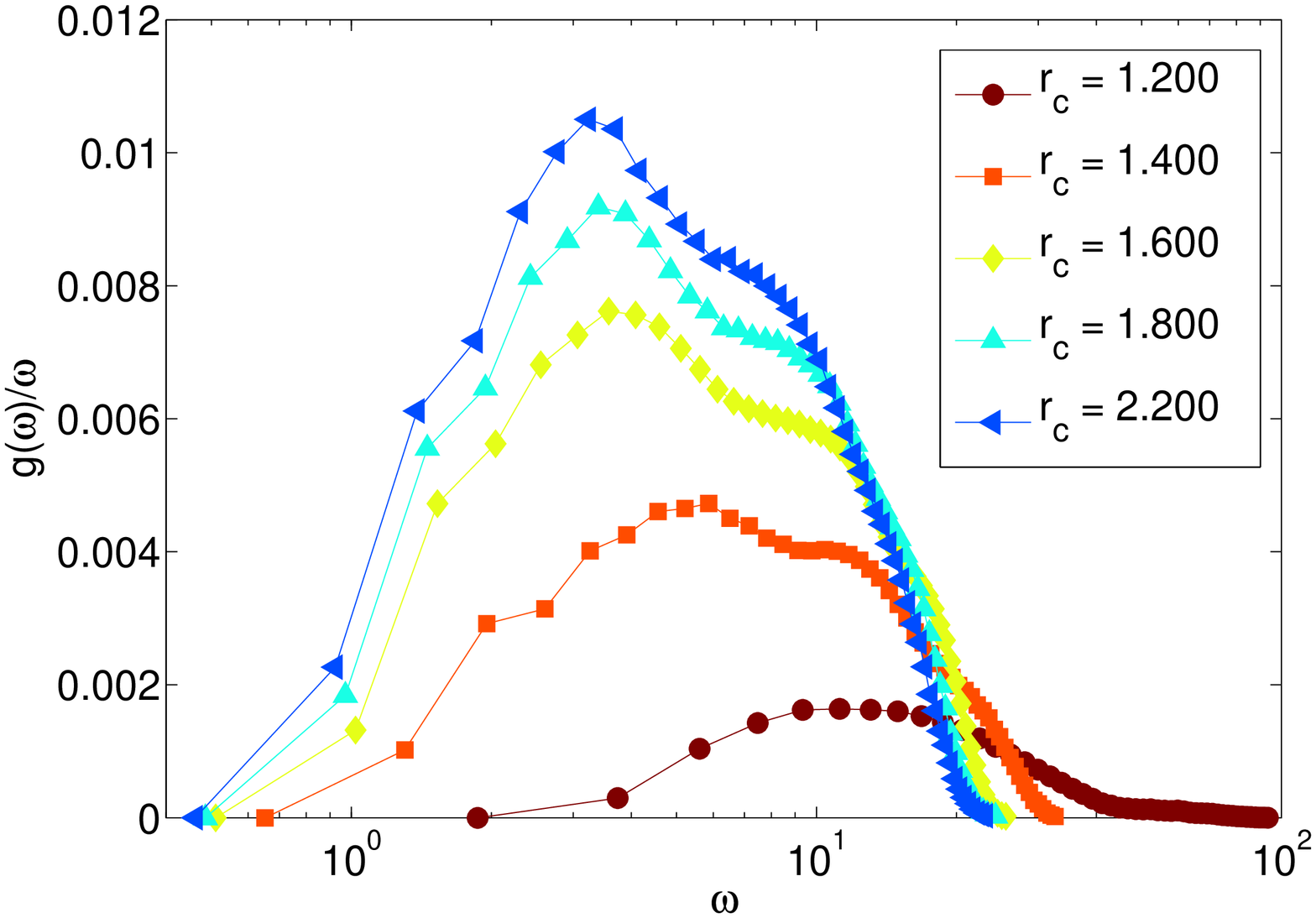}
\caption{Top panel: Variation of excess vibrational density of states (VDOS), 
$g(\omega)/\omega^2$ as a function of $\omega$ with cut-off range, $r_{co}$ 
for the model A in three dimensions. Notice the rapid change this quantity as 
a function of increasing cut-off range, $r_{co}$. Bottom panel: excess VDOS 
obtained for the same model in two dimensions. The variation with increasing
$r_{co}$ is really large.}
\label{fig:4}
\end{figure} 
Another thing to notice is that the variation of $\Gamma$ with the cut-off range of the 
inter-atomic potential in two and three dimensions are quite large. The value of 
$\Gamma$ changes almost $3.3$ times when we vary the cut-off from $1.2$ to $2.2$ for
the two dimensional system, while for the three dimensional system the change is 
approximately $1.6$ times. Hence, in both the systems we could see that as the 
cut-off increases the amount of plastic deformation that the system can withstand 
also increases dramatically before failure. This can be attributed to the variation 
in the number of neighbors of the atoms as we change the cut-off. Cut-off value of 
$1.2$ corresponds to the first peak in the radial distribution function for the 
system.
So in this case, the potential range is shorter than the first shell of the neighbors. 
Hence in this case, while the system is subjected to an external 
strain if any cavity forms which is larger than the first shell of neighbors it will 
not heal since there will be no attractive forces across it leading to abrupt failure. 
As we increase cut-off range more and more neighbors come in and we need a larger 
cavity to induce failure \cite{2011Dauchot} or the mechanism of failure may change 
completely.

To understand what actually changes in the system with increasing interaction 
range that leads to the observed brittle to ductile transition, we calculated  
the vibrational density of states (VDOS) to investigate a possible relationship 
between the mechanical properties of the system with that of the vibrational 
properties of the system as suggested in \cite{2011Dauchot, 2013Mizuno, 2014Goodrich}. 
The density of states were calculated from the eigenvalues of the Hessian matrix
obtained at the potential energy minimum of the system explored at that temperature.
The Hessian matrix is defined as 
$H_{ij}^{\alpha\beta} = \partial^2 U/\partial r_{i}^{\alpha}\partial r_{j}^{\beta}$
where $U$ is the potential energy of the system which is a function of the particle
coordinate $\vec{r}_{i}$ with $i$ indicates the particle index and $\alpha$ is
different components of the space dimension (x, y or z). The potential energy $U$
is first minimized with respect to the particle coordinates $\vec{r}_i$ using 
conjugate gradient methods and then used LAPACK routine to diagonalize the matrix
$H_{ij}^{\alpha \beta}$ which is $Nd \times Nd$ matrix, where $d$ is the spatial 
dimension.

The resulting VDOS $g\left ( \omega  \right )$ obtained as a function of $\omega$ 
is reduced to the form of $g\left ( \omega  \right )/\omega ^{d-1}$, where $d$ is the 
dimensionality, and the reduced VDOS is plotted in Fig. \ref{fig:4}. This is done
so to plot only the excess part of the density of states over Debye density of 
state of ideal solid where $g\left ( \omega  \right ) \sim \omega^{d-1}$. One can
clearly see in Fig. \ref{fig:4}, that the excess VDOS increases dramatically with 
increasing interaction range both in two and three dimensional model system. This
strongly suggests that excess vibrational density of state intimately connected the
mechanical properties of these systems as also suggested in Ref.\cite{2011Dauchot}. 
\begin{figure}[!h]
\includegraphics[width=0.45\textwidth]{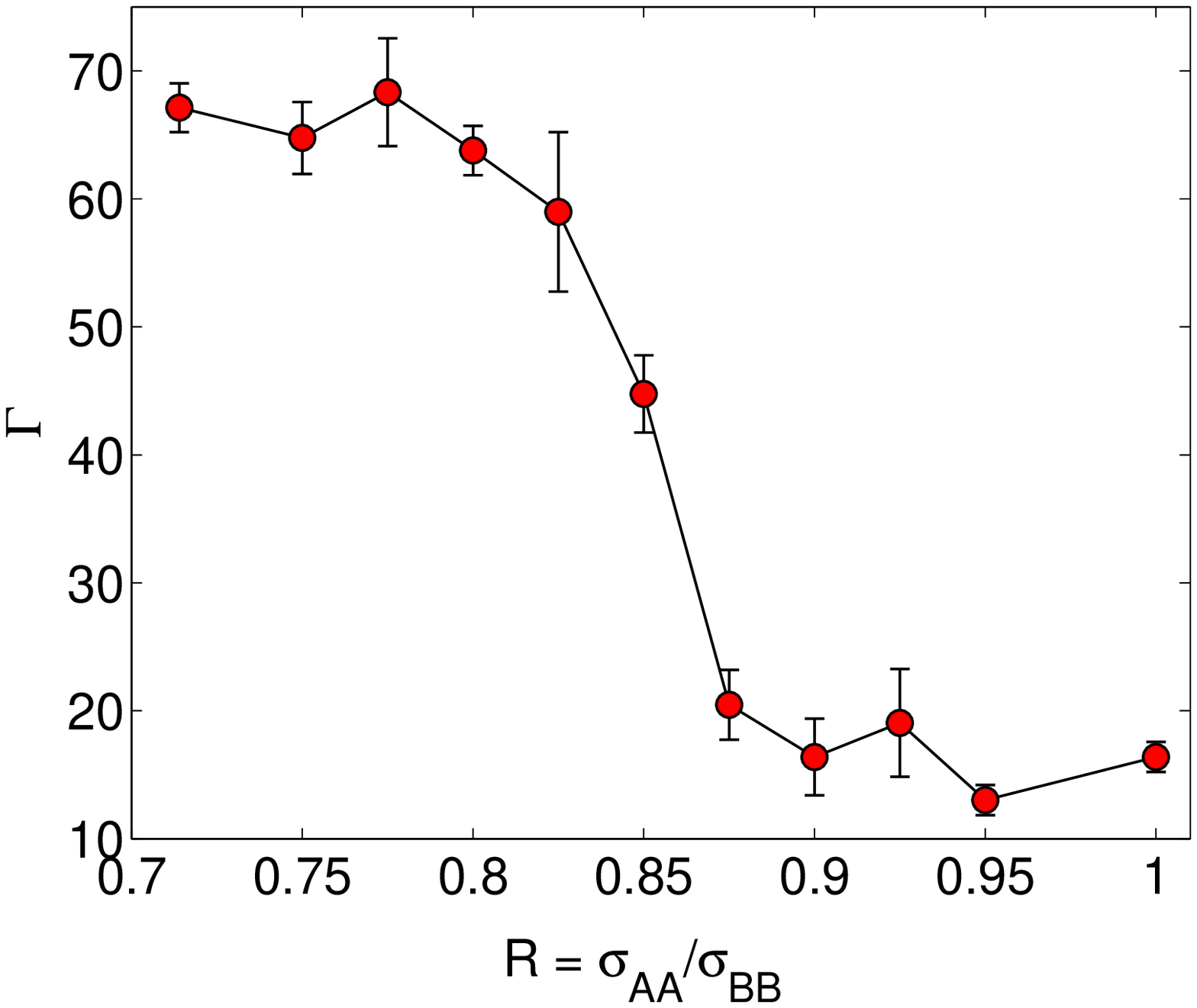}
\hskip -0.1cm
\includegraphics[width=0.47\textwidth]{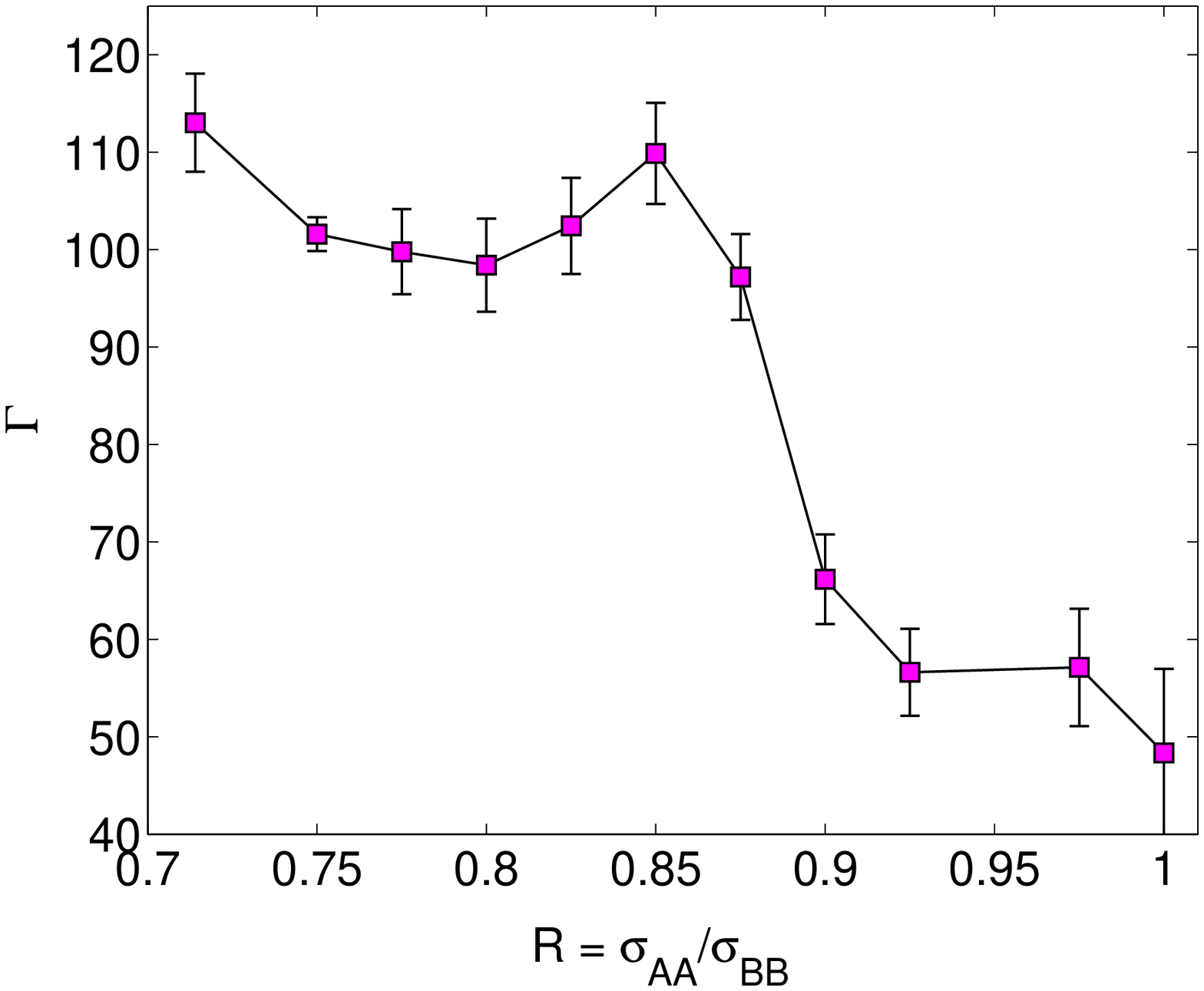}
\caption{Top panel: Variation of $\Gamma$ with size disorder characterized by 
$R = \sigma_{AA}/\sigma_{BB}$ (see text for details) for model A in two dimension. 
Notice the sharp change in $\Gamma$ at $R \sim 0.85$. Bottom panel: Similar 
analysis done for the same model in three dimensions. The sharp decrease in 
$\Gamma$ at $R \sim 0.90$ is clearly visible.}
\label{fig:3}
\end{figure}

The correlation between the excess VDOS and ductility of materials although suggest
interesting connection between the low energy vibrational states with plasticity
in the system but does not prove the unique dependence of ductility on excess 
VDOS. To understand whether excess VDOS can uniquely determine the ductility of
any material, we have studied the brittle-ductile transition across the amorphization
transition where the ratio of particle diameter of the binary glass forming liquids
is changed systematically from 1. This way one can go from crystal to glassy state 
by tuning the ratio of particle diameter in the model. Notice that as we will be 
going from crystal to a glassy state with varying the diameter ratio, the excess 
VDOS will increase very sharply as an ideal crystal will not have any excess VDOS 
and it will gradually increase as one goes to amorphous solids. This is an ideal 
set up to understand the role of excess density of state in brittle ductile
transition as the excess VDOS can be tuned very systematically.

\begin{figure*}
\centering
\includegraphics[width=1.0\textwidth]{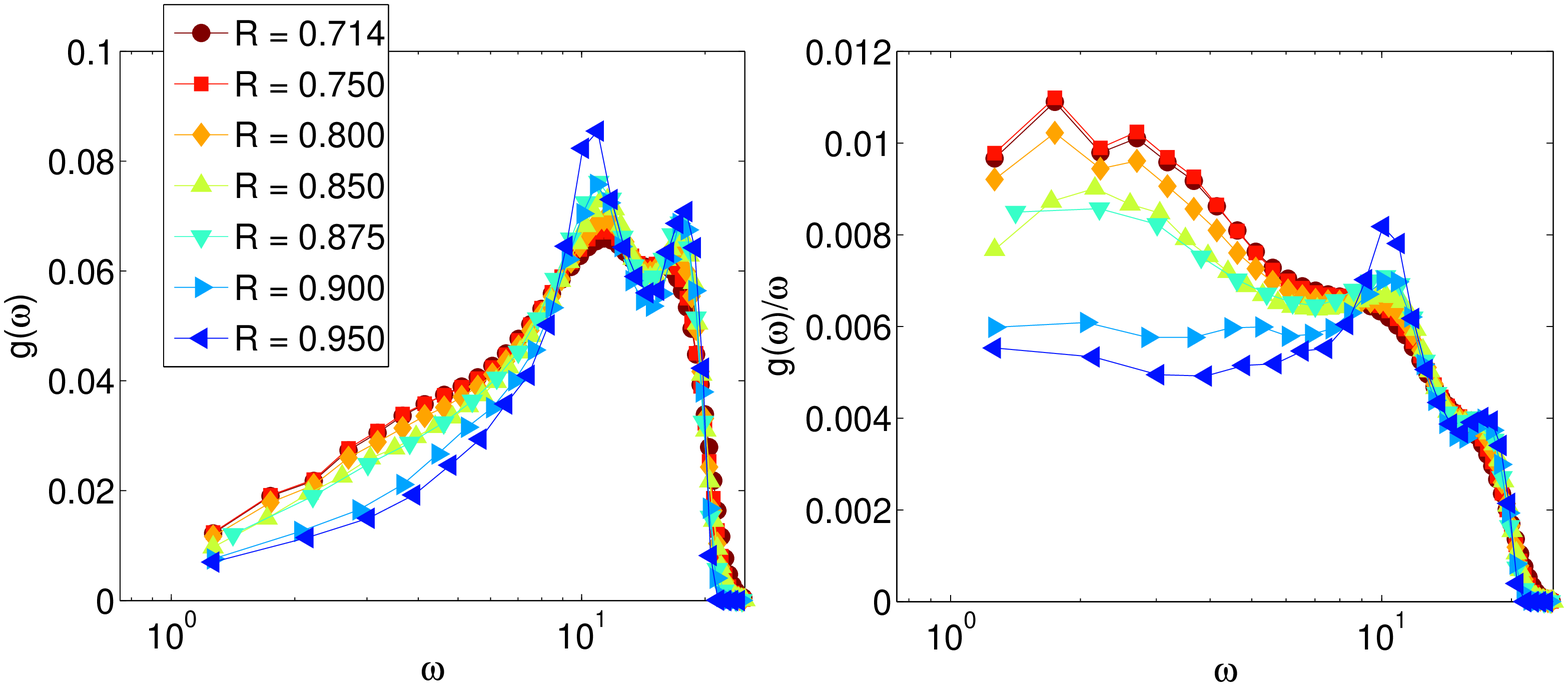}
\includegraphics[width=1.0\textwidth]{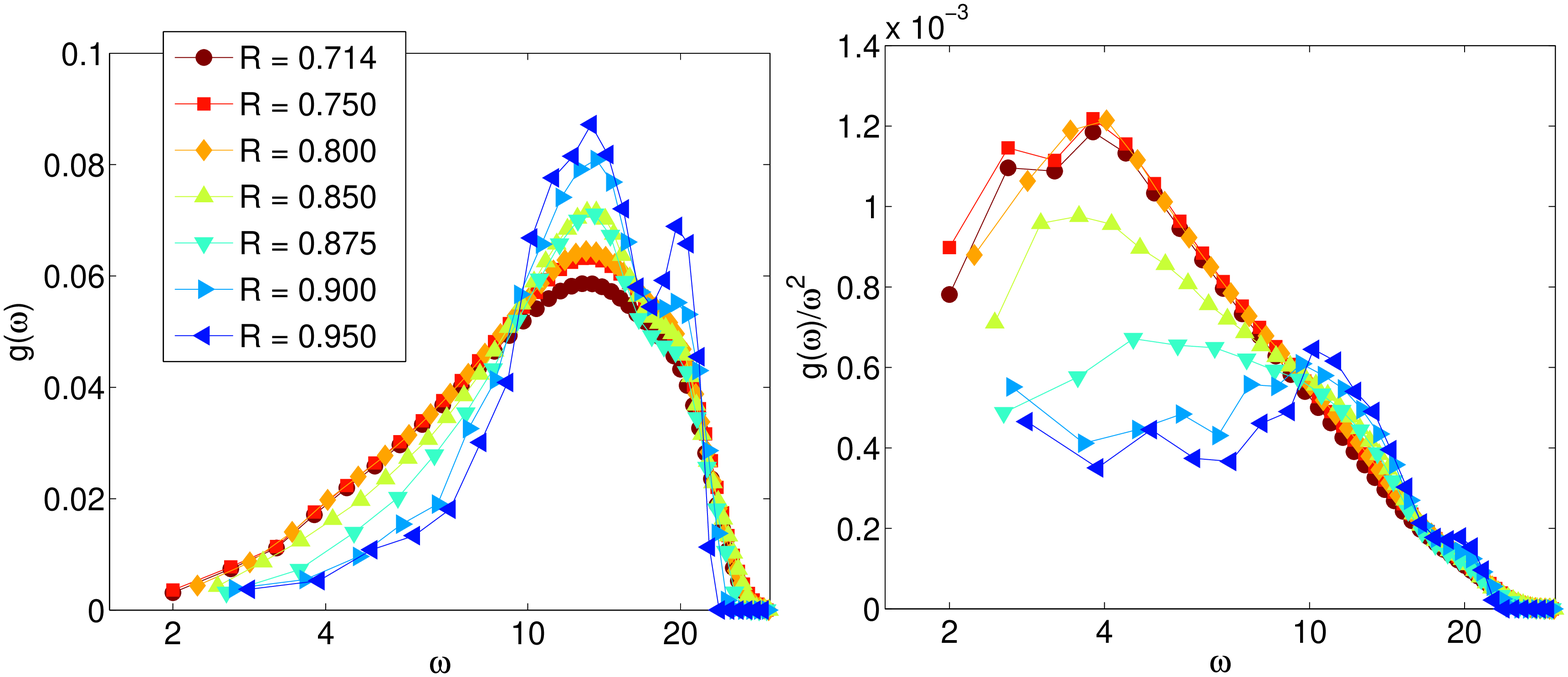}
\caption{Top left panel: Variation of VDOS as a function of $\omega$ with 
size disorder $R$ for model A in two spatial dimensions. Notice that 
$g(\omega)$ at small $\omega$ gains more weight as one goes from crystal 
to amorphous solids via the amorphization transition (see text for details) 
by decreasing $R$. 
Top right panel: $g(\omega)/\omega$ is plotted to clearly show the increase
in excess VDOS at small $\omega$. Bottom panels: Same analysis done for 
the model A in three dimensions.}
\label{fig:5}
\end{figure*}
In Fig. \ref{fig:3} we have shown the variation of $\Gamma$ with different values
of the ratio of particle diameter, $R = \sigma_{AA}/\sigma_{BB}$ for two and three 
dimensional systems respectively. $\sigma_{AA}$ is the diameter of the A type
particle and $\sigma_{BB}$ is the diameter of B type particle in a typical binary
mixture. Size ratio $R = 1$ corresponds to a mono-disperse system which will 
crystallize if the temperature is decreased below the freezing temperature. $R$ 
is varied from $1.0$ to the lowest size ratio of $0.714$ is similar to the size 
ratio used in the model system mentioned above. The interaction range is $r_c = 1.60$. 
In the top panel of Fig.\ref{fig:3}, the variation of $\Gamma$ is shown as a function
of $R$ for two dimensional system and one can see that $\Gamma$ remains more or less
independent of $R$ up to $R < 0.85$ and then sharply decreases at $R \simeq 0.85$ and
remains independent again up to $R = 1.0$. $\Gamma$ changes by a factor of $7.0$ in 
this model system. Similar behaviour is found for the three dimensional case with
$\Gamma$ changes by a factor of $2.0$. It is important to notice that the change of
$\Gamma$ as a function of $R$ in both dimensions, remains very sharp, suggesting a 
possible sharp transition from brittle to ductile materials at a critical dispersity
ratio $R = R_c \simeq 0.85$. 
\begin{figure*}[!ht]
\centering
\includegraphics[width=0.9\textwidth, angle =0]{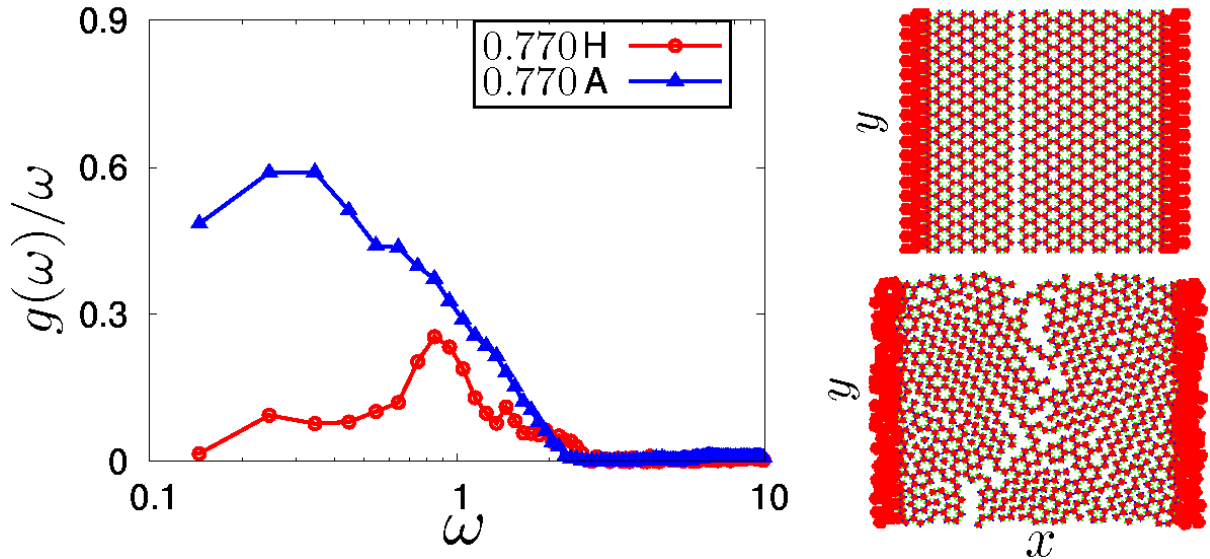}
\caption{Left panel: Variation of excess VDOS, $g(\omega)/\omega$ as a function of 
$\omega$ for the crystalline state (red circles) and the quenched amorphous state 
(blue triangles). Right top panel: Snapshot picture of the material in crystalline 
state under uniaxial strain along x-direction. Similar snapshot in the amorphous 
state (bottom right panel). }
\label{patchyVDOS}
\end{figure*}

Now to prove the unique connection between the excess VDOS and the ductility we have
calculated the VDOS for all these systems which are plotted in Fig.\ref{fig:5}. In top
left panel of Fig.\ref{fig:5}, we have shown the VDOS for the two dimensional model
system for different dispersity ratio $R$ and in right panel of the same figure 
the excess VDOS is plotted. One can clearly see that for $R>0.85$, the excess VDOS
is almost negligible as $g(\omega)/\omega$ is almost independent of $\omega$ in the small $\omega$ limit and increases sharply once $R<0.85$ and becomes almost
insensitive to change in $R$ for lower values. Similar behaviour is observed for the
three dimensional case also. 

These observation clearly establishes the one to one 
connection between the excess VDOS and the ductility of a material irrespective of
whether it is a crystalline solid or an amorphous solids. As mentioned earlier, in 
\cite{2014Goodrich}, it was shown that mechanical properties of a slightly 
defected crystal with a very small density of vacancies is more close to a jammed 
amorphous solid than to the perfect crystal. They also pointed out that vibrational 
density of states of a solid with even a very high degree of crystallinity resembles 
more closely to that of the amorphous jammed solid. A sharp transition in mechanical 
and vibrational properties is also suggested in that work. Our work here 
similarly suggest a possible sharp brittle to ductile transition as one goes from 
crystalline to amorphous solid by changing dispersity parameter, $R$. It further points 
out a one-to-one correspondence between the mechanical and vibrational properties of 
solids.

A very similar connection between the observed behaviour in our work and behaviour 
observed in fiber bundle model \cite{fibreBundle10PHC_RMP,fibreBundle12KHKBC_RMP} 
with increasing dispersity in the fiber strength is
worth mentioning. The fiber bundle model is a simplistic model to understand fracture
in materials. In this model, two plates are connected by fiber bundles and then the 
plates are loaded to mimic the tension experiments
\cite{fibreBundle10PHC_RMP,fibreBundle12KHKBC_RMP}. The solid will then be modeled
via the fiber in between the plates. Now if one takes the strength of these fibers
to be same then it will model a homogeneous solid, but if one takes the strength of
the fiber to be heterogeneous, then it will be closely mimicking the behaviour of an 
amorphous solids. In \cite{15purusattamRay}, one such model is studied with the 
strength of fibers taken from a Gaussian distribution of width $\delta$. This disorder
parameter $\delta$, was then varied and the mechanical properties of the solid is 
studied. It was shown that as one increases the disorder parameter $\delta$, the solid
modeled by the fiber bundle model shows a brittle to ductile transition at a critical
disorder strength $\delta = \delta_c$. Our results seems to strongly suggest such a
transition in realistic model materials \cite{97ZREVPRL}, thereby providing a nice 
model system to test the prediction of these fiber bundle models for further 
improvement of these minimalistic models \cite{15PBDgK}. 

\subsection{Model B:}
So far we have tested the connection between the excess VDOS and the ductility for 
system where the interaction potential between two particles is isotropic. To test
whether the same connection still holds for other model interaction potentials with
anisotropic interaction, we have performed the similar analysis for a patchy colloid
model described in the model and simulation details section before. This model at 
different temperature and density shows many interesting phases (see \cite{15MKS} 
for further details and the phase diagram), {\it e.g} triangular, honeycomb, square
lattice and their corresponding phase coexistence. For a large portion of the phase diagram, this model also shows amorphous structure. In this case, an amorphous solid may be formed by quenching relatively rapidly into a region of the phase diagram where there are many competing crystalline states. The solid, instead of selecting between these almost degenerate free energy minima corresponding to very different structures, chooses to remain amorphous or forms microscopic crystallites each with wildly varying local coordination. 

We perform loading experiments using the protocol described in section~\ref{sec:model} for this model solid at two representative states.  Both the solids are at a number density $\rho = 0.77$, where the system at 
low temperature forms a honeycomb lattice. The first state is crystalline with two dimensional honeycomb order obtained by arranging particles in an initial crystalline order and equilibrating the structure at the chosen density and temperature. The second state, at the same density, is amorphous and is formed by a temperature quench from the high temperature liquid. Note that this amorphous structure is clearly in a state of dynamic arrest. 

For the crystalline state one has no excess density of state over the Debye theory and for the amorphous state there is an excess of states as shown in Fig.\ref{patchyVDOS}a. Now, according to our 
previous observations, the state with excess VDOS should be more ductile than
the one with less excess VDOS. It is indeed true that the amorphous state with 
excess VDOS can withstand longer tensile strain and bears $\Gamma \simeq 32$, while the
pure crystalline state with less VDOS breaks like a brittle materials with 
$\Gamma \simeq 1.5$. Fig.\ref{patchyVDOS}b shows one instance
of fracture each in the crystalline and amorphous states obtained for the patchy 
colloidal system with different preparing protocol. The crystalline state shows a 
sharp fracture while the amorphous state seems to show more heterogeneous fracture
profile. The increase in low
frequency vibrational states for the amorphous structure is quite dramatic. 
Thus we believe that the excess VDOS uniquely captures the mechanical behaviour 
of a solid with any form of inter-atomic interaction potential and irrespective of 
whether it is in a crystalline or amorphous state.         

\section{\label{sec:conclusions}Conclusions}
In conclusion, we have done extensive MD simulation of different 
glass forming liquids with both isotropic and anisotropic pairwise interactions to 
understand the microscopic origin of ductility in amorphous solids. There has been many attempts to link macroscopic mechanical properties of solids such failure mechanism with microscopic ``atomic'' interactions. For example, extensive calculations and analysis of known experimental data for a number of metallic and non-metallic solids by Rice and Thomson \cite{74riceThomson} shows that ductility in solids is related to the ability of a crack tip to nucleate dislocations. Such dislocations nucleated at crack tips blunt the crack and produce ductile behaviour. A confirmation of this scenario has been found recently in the computer simulations \cite{99HartmaierGumbsch,marderBook} where dislocation tangles surrounding cracks in ductile materials is readily observed in large scale computer simulations. 

In this paper, on the other hand we find a different kind of correspondence. We observe that
ductility of a material is intrinsically connected to excess density of vibrational
states over the Debye density of state of an ideal solid. Findings of Goodrich 
{\it et.al.} \cite{2014Goodrich}, clearly suggest that excess vibrational density 
of state is a unique characteristic of the material which strongly connects to the 
mechanical property of the system. They also showed that solid with not perfect 
but large crystalline order show properties closely matching with that of an
amorphous jammed solid than ideal crystal with defects. The vibrational density 
of state of these systems becomes very close to that of an amorphous solids. Here
we showed that excess vibrational density of states uniquely determines the
ductility of an amorphous solids under external tensile load. We also showed that
there exists a sharp brittle to ductile transition as one drives the system from 
crystalline to amorphous state via amorphization transition which is in close 
agreement with the predictions of minimalistic fiber bundle model 
\cite{15purusattamRay}. It will be 
nice in future to understand in details this transition by doing extensive 
finite size scaling analysis. Study along these directions are in progress and 
will be published else where \cite{15PBDgK} Our results on the patchy colloidal 
model also confirms that the reported results are rather generic for any glass 
forming liquids. 

Is there a connection between VDOS and the ability of the solid to produce dislocations? While a complete answer to this very interesting question has not been found, there are intriguing possibilities. In \cite{15saswatiSM} it was shown that probability of formation of defects in crystalline solids depends on the amplitude of certain non-affine displacement fluctuations  \cite{15saswatiPRE} which act as precursors. Non-affine defect precursors also cause instabilities in the crystal leading to mode softening which may, in turn enhance the VDOS in the small frequency regime. We thus feel that our finding of direct connection between excess density of vibrational states and plasticity in amorphous solid as well as 
crystalline solids will help us better understand plasticity in these materials
within an unified theoretical framework. 

%

\end{document}